\documentclass[aps,amsfonts,prl,twocolumn,showpacs]{revtex4-1}

\usepackage{subfigure}
\usepackage{wrapfig}
\usepackage{textcomp}
\usepackage{graphicx}
\usepackage{amssymb}
\usepackage{amsmath}
\usepackage{bm}
\usepackage{amsmath, amsthm, amssymb}
\usepackage{hyperref}

\def\beq{\begin{equation}}
\def\eeq{\end{equation}}
\def\bea{\begin{eqnarray}}
\def\eea{\end{eqnarray}}

\begin{document}

\title{Quantum quasicrystals of spin-orbit coupled dipolar bosons}

\author{Sarang Gopalakrishnan$^1$, Ivar Martin$^2$, and Eugene A. Demler$^1$}

\affiliation{$^1$Department of Physics, Harvard University, Cambridge MA 02138, USA \\ 
$^2$ Materials Science Division, Argonne National Laboratory, Argonne, Illinois 60439, USA}

 \date{\today} 

\begin{abstract}
We study quasi-two-dimensional dipolar Bose gases in which the bosons experience a Rashba spin-orbit coupling. We show that the degenerate dispersion minimum due to the spin-orbit coupling, combined with the long-range dipolar interaction, can stabilize a number of quantum crystalline and quasicrystalline ground states. Coupling the bosons to a fermionic species can further stabilize these phases. We estimate that the crystalline and quasicrystalline phases should be detectable in realistic dipolar condensates, e.g., dysprosium, and discuss their symmetries and excitations. 
\end{abstract}

\maketitle

Quasicrystals are exotic spatially ordered states of matter that have no periodic crystal lattice~\cite{primer, shechtman, *steinhardt:prl}. Quasicrystalline states have been observed in metallic alloys~\cite{tsai:review, steinhardt:science} and colloids~\cite{lifshitz2011}, and engineered in  photonic crystals~\cite{lifshitz2007}. Quasicrystals differ dramatically from crystals in their mechanical~\cite{primer, lifshitz2007} and electronic properties~\cite{primer}; like most crystals, however, the existing quasicrystals are classical, in the sense that quantum statistics does not affect the crystalline order. Thus, the unconventional excitations, defects, and melting transitions~\cite{berg09, *agterberg08, *agterberg11, sg:disclin} that arise when translational order is intertwined with Bose condensation (as in liquid-crystalline states of paired electrons~\cite{ff, lo, radz, *radz2, berg09, *agterberg08, *agterberg11}) cannot be accessed with existing quasicrystals. Even classical quasicrystals have unconventional excitations (e.g., phasons~\cite{primer} and imperfect dislocations~\cite{kleman1, *kleman2}); their quantum analogs should therefore be unusually rich in such excitations and associated phenomena~\footnote{We emphasize that our work concerns \emph{emergent} quasicrystalline order, as opposed to electronic motion in externally imposed quasicrystalline potentials, such as (e.g.) incommensurate optical lattices.}. While quantum (quasi)crystals are difficult to realize in solid state systems, ultracold atomic gases offer naturally quantum-degenerate, tunable platforms for studying the interplay between (quasi)crystallinity and Bose condensation. Indeed, various proposals for studying the resulting ``supersolid'' and ``supersmectic'' phases exist~\cite{goral, lewenstein:roton, buechler, stamper, cherng, sg:np, *sg:pra, baumann2010, pohl2010, pupillo, *pupillo2, *pupillo3, sds:bfsolid, stringari}. However, these proposals chiefly consider two-dimensional stripes and triangular-lattice crystals.

\begin{figure}[b]
\begin{center}
\includegraphics{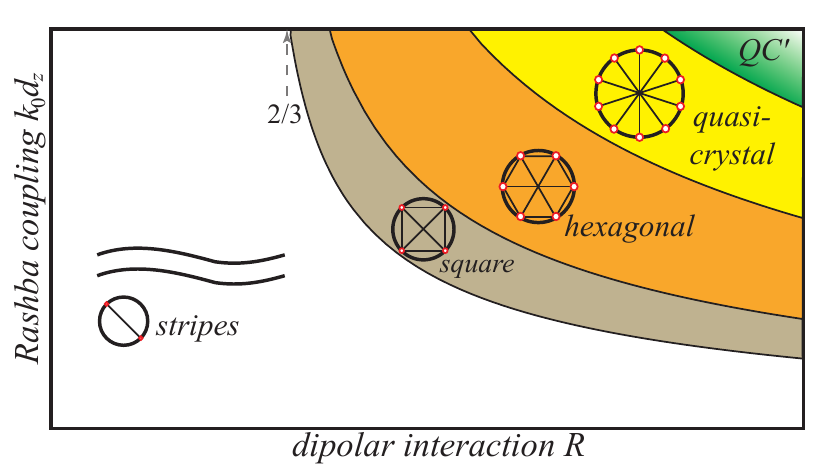}
\caption{Ground state phase diagram of two-dimensional Rashba-coupled dipolar bosons. The horizontal axis is the parameter $\mathcal{R}$---defined in the text---which measures the ratio of the dipolar interaction to the contact interaction. The vertical axis is $k_0 d_z$, where $k_0$ is the Rashba coupling strength and $d_z$ is the transverse confinement. In the region marked \emph{QC'} the system exhibits quasicrystalline phases involving $\geq 7$ pairs of momenta.}
\label{fig:phasediag}
\end{center}
\end{figure}

In this Letter, we present an approach for realizing more general crystalline and quasicrystalline states of ultracold bosons (Fig.~\ref{fig:phasediag}). We consider bosons subject to a Rashba spin-orbit coupling, which gives rise to a single-particle dispersion minimized on a circle in momentum space~\cite{campbell, galitski, zhai, sg:lamacraft}. This momentum-space circle sets the lattice spacing; (quasi-)crystalline phases correspond to condensation at a discrete set of momenta on it. The Rashba coupling alone does not generate nontrivial crystals: Rashba-coupled bosons~\cite{mondragon, galitski, zhai, *zhai2, generic, sg:lamacraft, ozawa2, barnett, stringari, wureview, goldmanreview} exhibit a striped state, involving condensation at two opposite momenta, and a spatially homogeneous state, involving condensation at a single momentum, as well as possible uncondensed states~\cite{sedrakyan}. However, as we show here, adding a second length-scale via dipolar interactions can stabilize nontrivial crystalline and quasicrystalline states (Fig.~\ref{fig:phasediag}). This stabilization takes place through a mechanism that differs from the conventional accounts of quasicrystalline ordering~\cite{bak, troian}. All the spatially ordered states we find are inherently quantum-mechanical, in that the relative $U(1)$ phases of their momentum-components affect the crystallinity; we show that they also exhibit additional, distinctively quantum-mechanical ``phason'' excitations. 

We estimate that these crystalline and quasicrystalline states are achievable in realistic experiments and can, in addition, be easily detected via time-of-flight imaging, as their momentum distributions are sharply peaked at reciprocal lattice vectors. Although we focus on ultracold atomic realizations, the results of the present work might also apply to certain magnetic systems, such as MnSi~\cite{turlakov, *pfleiderer, *bishop-rkky}, in which the spin-wave dispersion has a circular minimum. 


\emph{Model}. The model we consider here comprises two-dimensional dipolar bosons, subject to a Rashba spin-orbit coupling. The Hamiltonian can be written as 

\beq\label{Ham}
\mathcal{H} = H_0 + H_{\mathrm{int.}}
\eeq
where 

\beq
H_0 = \frac{1}{2m} \sum_{\mathbf{k}, \alpha\beta} \phi^\dagger_\alpha(\mathbf{k}) \left(\mathbf{k}^2 \delta_{\alpha\beta} + \lambda \mathbf{(k \times \sigma_{\alpha\beta}) \cdot \hat{z}} \right) \phi_\beta(\mathbf{k})
\eeq
is the single-particle Hamiltonian, with a circular dispersion minimum at $k_0 \equiv \lambda/2$. Various schemes exist (see, e.g., Ref.~\cite{campbell}) for realizing $H_0$ using multiple Raman-coupled internal states; such schemes should be possible to realize in strongly dipolar Bose condensates such as those in dysprosium~\cite{dybec, levzhai} and erbium~\cite{erbiumbec}, whose ground states have many Zeeman sublevels~\footnote{B.L. Lev (private communication)}. 

\begin{figure}[htbp]
\begin{center}
\includegraphics{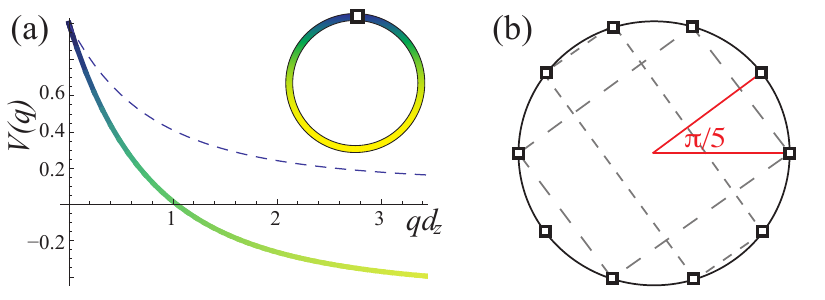}
\caption{(a)~Interaction potential $U(q)$ as a function of momentum $q$ for $R = 0.6$ (dashed line) and $R = 1$ (thick line). Inset shows the interaction potential in momentum space due to a single condensate at $k_0 \mathbf{\hat{y}}$ in the $R = 1$ case; the color-coding matches that in the main panel, and indicates that the interaction energy is attractive for a condensate sufficiently far away in momentum space. (b)~Ten-component (pentagonal) quasicrystal pictured in momentum space; dashed squares show typical momentum combinations that are coupled by interactions.}
\label{interactions}
\end{center}
\end{figure}

The interaction, $H_{\mathrm{int}}$, including both contact and dipolar terms, is assumed to be a density-density interaction. This assumption generally holds for the contact interaction; the dipolar interaction is also chiefly density-density if the states used to give the spin-orbit coupling are large-spin states, such as $m_F = 7, 8$ in dysprosium. (Wilson et al.~\cite{wilson} have considered the opposite limit of strongly spin-dependent dipolar interactions.) The full interaction takes the momentum-space form 

\beq
H_{\mathrm{int}} = \int d^2 k \rho_{\mathbf{k}} \rho_{\mathbf{-k}} U(k)
\eeq
where $\rho_{\mathbf{k}} \equiv \int d^2x e^{i \mathbf{k \cdot x}} (\phi^\dagger_+ (\mathbf{x})\phi_+ (\mathbf{x}) + \phi^\dagger_- (\mathbf{x}) \phi_- (\mathbf{x}))$ and $U(k)$ is given by~\cite{fischer, mehrtash0, *mehrtash1}:

\beq
U(k) = U(0) \left[1 - R k d_z w\left(\frac{k d_z}{\sqrt{2}} \right) \right]
\eeq 
where $d_z$ is the confinement strength in the $z$ direction; $U(0)$ is the overall strength; $R \equiv (3/2) \sqrt{\pi/2}/(1 + g_0/g_{d})$, where $g_0$ and $g_d$ are the contact and dipolar scattering lengths respectively; and $w(z) \equiv \exp(z^2) \mathrm{erfc}(z)$. For $R \rightarrow 0$ this interaction becomes purely contact; for $R > 2/3$, the dipolar interaction overcomes the contact interaction, and $U(k)$ changes sign at large $k$ [Fig.~\ref{interactions} (a)]. This regime, in which we find nontrivial ground states, is naturally achieved in dysprosium~\cite{dybec}, and can be engineered even for less dipolar species such as chromium, by tuning the contact interaction to zero via a Feshbach resonance. 

\emph{Mean-field analysis}. We now turn to a mean-field analysis of Eq.~\eqref{Ham}. We make the following ansatz for a crystalline state with paired momenta:

\bea
\phi(\mathbf{x}) & = & \sqrt{ \frac{n}{2M}} \sum_{i = 1}^M \bigg[ e^{i (\mathbf{k}_i \mathbf{\cdot x} + \alpha^+_i)} 
\left( \begin{array}{c} 1 \\ e^{i \theta_i} \end{array} \right) \nonumber \\ & & \qquad \quad + e^{-i (\mathbf{k}_i \mathbf{\cdot x} + \alpha^-_i)} 
\left( \begin{array}{c} 1 \\ -e^{i \theta_i} \end{array} \right)\bigg],
\eea
where $n$ is the total density; $\theta_i$ is the angle between $\mathbf{k}_i$ and the $x$ axis; and $M$ is the number of density-waves composing the (quasi-)crystal. The restriction to equal-weight states is justified in the Supplemental Material; the assumption of paired momenta is justified below (and also by renormalization arguments~\cite{sg:lamacraft}). For such states, the interaction energy density per particle is given by

\beq\label{mf}
E = U(0) + \frac{1}{(2M)^2} \sum_{i \neq j, \pm} U(|\mathbf{k}_i \pm \mathbf{k}_j|) F^\pm_{ij} 
\eeq
where $F^\pm_{ij} \equiv 2 + \cos(\alpha_{ij}) + \cos(\alpha_{ij} + 2 \theta_{ij}) \mp \cos(\alpha_{ij} + \theta_{ij}) \mp \cos(\theta_{ij})$; $\alpha_{ij} \equiv (\alpha^+_i - \alpha^-_i) - (\alpha^+_j - \alpha^-_j)$; and $\theta_{ij}$ is the angle between components $i$ and $j$ on the dispersion minimum (chosen to be $\leq \pi/2$).  
%
The $\alpha$-dependent terms arise because of scattering processes of the form $\phi^\dagger_{\mathbf{k}_1, \alpha}\phi^\dagger_{-\mathbf{k}_1, \beta} \phi_{\mathbf{k}_2, \alpha} \phi_{-\mathbf{k}_2, \beta}$. In states where the momenta are not paired, these processes do not exist; thus, the mean-field energy of such states is generically higher, justifying our neglect. 

\begin{figure}[t]
\begin{center}
\includegraphics{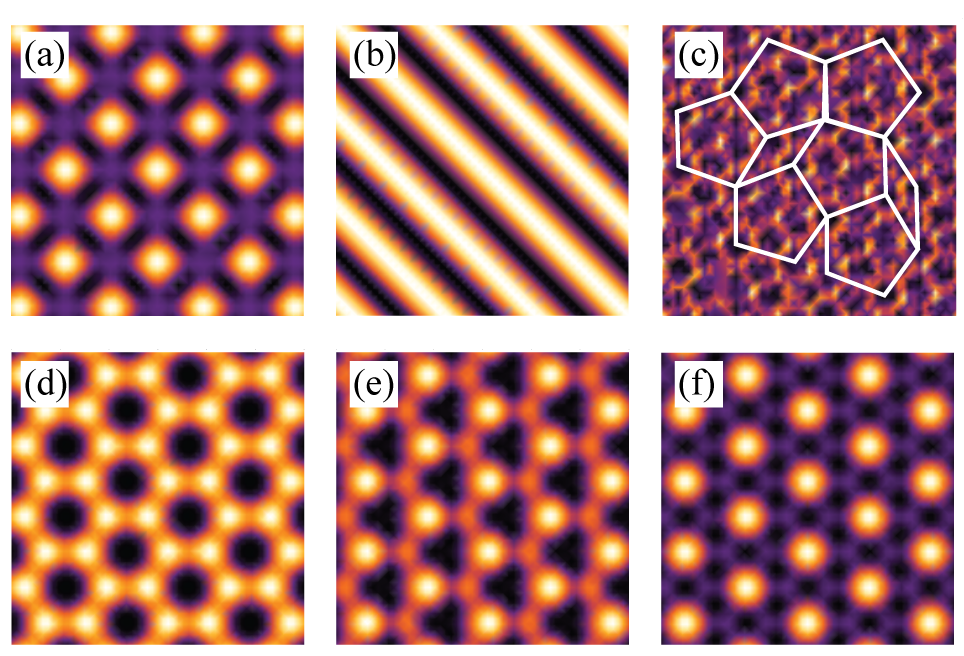}
\caption{Real-space spin and total density plots for various ordered states. (a)~Density of the spin-up component of the ``square-lattice'' phase. (b)~Total density profile of the square-lattice phase. (c)~Total density profile for the quasicrystalline phase; pentagons are guides to the eye. (d)-(f)~Total density profile of the hexagonal phase, as a function of the relative phase $\Xi_{\mathbf{k}}$ [Eq. (10)] of any one of the three condensate pairs. Varying this phase smoothly from $0$ [panel (d)] through $\pi$ [panel (f)] corresponds to an overall translation plus a phason.}
\label{prettypics}
\end{center}
\end{figure}

We now consider the limiting cases. The simplest limit is the pure contact interaction limit $R = 0$, for which $U(q) \equiv U(0)$ is constant. In this limit, for a generic $M$-component state

\beq\label{contact}
E = U(0)\!\! \left[1 + \frac{1}{M^2}\sum_{i < j} \{1 + \cos(\alpha_{ij} + \theta_{ij}) \cos\theta_{ij}\}\right] 
\eeq
The second term is always positive, except for $M = 1$ (in which case it is absent); therefore, the ground state is always a stripe, with $E = U(0)$. Provided that $U(q) > 0$ for all $q$, this situation obtains, and the ground state remains a stripe.

We now turn to the dipole-dominated limit in which $U(q) < 0$ for large $q$; in this limit, the parameter tuning between different phases is $k_0 d_z$, the ratio of the spin-orbit coupling scale to the transverse confinement scale. For $k_0 d_z \ll 1$, the interaction potential takes the form $U(k) \sim U(0) (1 - R k_0 d_z) > 0$, so the additional terms in Eq.~\eqref{mf} are positive, and a stripe is still the lowest-energy state. In the opposite limit $k_0 d_z \gg 1$, however, $U(k)$ is \emph{attractive} between points that are sufficiently far apart on the dispersion minimum, thus favoring multi-component crystals and quasicrystals. 

Directly minimizing the interaction energy Eq.~\eqref{mf} in this regime yields the phase diagram in Fig.~\ref{fig:phasediag}; we see that increasing $k_0 d_z$ gives rise to crystals and quasicrystals with increasingly many momentum components. This feature can be understood heuristically as follows. The interaction potential Fig.~\ref{interactions}(a) is negative and approximately constant for $k d_z \gg 1$; thus, one can think of it as having a repulsive core at $k \alt 1/d_z$ and a constant attractive tail of strength $U_\infty$ at $k \agt 1/d_z$. Thus, the interaction energy is minimized by fitting as many condensates as possible around the dispersion minimum at a momentum spacing $\geq 1/d_z$; in the Supplemental Material we show that the energy of a generic $M$-component state is then

\beq\label{heuristic}
E = U(0) - \frac{|U_\infty|}{2M} \left[\! (M - 1) + \!\!\!\!\!\!\!\!\! \sum_{n \leq (M - 1)/2} \!\!\!\!\!\!\!\! 2 \cos(n \pi/M) \!\right],
\eeq%
which is minimized by increasing $M$ until it is $\sim k_0 d_z$. The resulting state can be regarded as a Wigner crystal on the dispersion ring, in addition to being a real-space (quasi)crystal. In general, such an $M$-component arrangement does not correspond to any periodic crystal lattice; however, it \emph{does} have sharp Bragg peaks, by construction, and is therefore a quasicrystal~\cite{primer}.

We now return to the results found by minimizing Eq.~\eqref{mf} and plotted in Fig.~\ref{fig:phasediag}, and note two features missed by the heuristic analysis. 
(1)~We find no asymmetric crystals, i.e., all states we have found involve condensation at an evenly spaced set of points on the dispersion minimum. (2)~States with even $M$ appear less stable than those with odd $M$; thus, for example, the square lattice ($M = 2$) occupies less of the phase diagram than the odd-$M$ states, and states with $M = 4, 6$ are always higher in energy than the nearest odd-$M$ state. This can be attributed to the fact that such states have momentum-components separated by $\pi/2$, for which the $\theta$-dependent terms in Eq.~\eqref{mf} vanish. 

We also note that, in the above discussion, we have labeled the phases by their pattern of spin-up (or spin-down) densities. Thus, for example, the phase we denote as a ``stripe'' is actually a spin-density wave with uniform total density~\cite{zhai}. Furthermore, our square-lattice phase has only a unidirectional density wave in the total density, as the spin-up and spin-down components form square lattices that are mutually out of phase [Fig.~\ref{prettypics}(a), (b)]. 

\emph{Stability against collapse}. We now discuss the stability of the mean-field phases against collapse, which is known to occur for dipolar Bose gases in certain regimes~\cite{pfau, fischer, dybec}. First, we see from Eq.~\eqref{heuristic} (which overestimates the attractive part of the interaction energy and is thus a lower bound) that the total interaction energy is always positive, even for pure dipolar interactions ($U_\infty = -0.3 U(0)$); thus, all configurations on the dispersion minimum are stable against global collapse. One can also check that there are no soft modes for $k \neq k_0$. Adapting Ref.~\cite{fischer} (see Supplemental Material) we find that this stability criterion takes the form 

\beq\label{stability}
n \leq 1/[U(0) m d_z^2] \simeq k_0^2 /[m U(0)].
\eeq
In terms of standard experimental parameters, the latter expression can be rewritten as the criterion that $n / k_0^2 \alt 1/ (k_0 a_d)$, where $a_d$ is the scattering length associated with the dipolar interaction. In general, $k_0 a_d \ll 1$, so Eq.~\eqref{stability} is satisfied for realistic densities ($n/k_0^2 \simeq n \lambda_R^2 \alt 1$, where $\lambda_R$ is an optical wavelength). This contrasts with previous proposals for realizing crystallinity through roton softening in dipolar gases~\cite{cherng, fischer}. The crucial difference between the two situations is that, in Refs.~\cite{cherng, fischer}, the dipolar interactions must be strong enough to overcome the kinetic energy if a crystal is to form, whereas, in the present case, the Rashba coupling quenches the kinetic energy on a momentum-space circle, so that \emph{some} kind of crystal forms even for very weak interactions. Therefore, in the present case, the kinetic energy \emph{away} from the dispersion minimum can be much larger than the interaction energy, thus ensuring at least the \emph{local} stability of the ground state.

\emph{Symmetries and gapless modes}. We now briefly enumerate the symmetries and thus the gapless modes of the $M$-component states. First, these states are symmetric under overall rotations; the corresponding long-wavelength orientational fluctuations suppress stripe ordering~\cite{zhai2} but not crystallinity. The other symmetries can be understood as combinations of the $2M$ condensate phases of an $M$-component crystal. These are locked in the combinations $\alpha_{ij}$ discussed above; we now discuss the significance of these combinations. 

Near a circle in momentum space, kinematic constraints~\cite{shankar} restrict two-particle interactions to be of two kinds: namely, forward-scattering processes of the form $(\phi^\dagger_{\mathbf{k}} \phi_{\mathbf{k}})(\phi^\dagger_{\mathbf{q}} \phi_{\mathbf{q}})$, and ``Cooper-channel'' processes of the form $\phi^\dagger_{\mathbf{k}} \phi^\dagger_{\mathbf{-k}} \phi_{\mathbf{q}} \phi_{\mathbf{-q}}$. The forward-scattering processes are invariant under the independent rephasing of each condensate; however, the Cooper-channel processes lock certain combinations of the condensate phases, namely the $\alpha_{ij}$. These processes are invariant under either (a)~a global change of phase, or (b) the joint transformations 
\beq
\phi_{\mathbf{k}} \rightarrow \phi_{\mathbf{k}} e^{i\Xi_{\mathbf{k}}}, \phi_{\mathbf{-k}} \rightarrow \phi_{\mathbf{-k}} e^{-i\Xi_{\mathbf{k}}}.
\eeq
Thus, there are $M + 1$ such modes for an $M$-component state. The one symmetry of type (a) is the overall $U(1)$ symmetry of the condensate, which gives rise to the superfluid stiffness. The $M$ remaining symmetries of type (b) correspond to sliding any one of the density waves comprising the (quasi-)crystal while leaving the others fixed. Two linear combinations of these generate rigid translations of the entire spatial structure and correspond to phonons; these combinations involve choosing $\Xi_\mathbf{k} \sim \mathbf{k \cdot G}$ for some $\mathbf{G}$. Note that for a standard Bravais-lattice crystal ($M = 2$), all type (b) transformations are phonons. However, for $M > 2$ there are $M - 2$ further symmetries; these are associated with excitations known as ``phasons''~\cite{primer, bak}. Phasons correspond to continuous internal rearrangements of a crystal that do not change its energy; a particularly intuitive example is afforded by the $M = 3$ hexagonal state. In this state, there is a single phason, pictured in Fig.~\ref{prettypics}(d)-(f); this excitation consists of continuously changing the density imbalance between the A and B sublattices of the honeycomb lattice, which leaves the interaction energy invariant. Note that, unlike the translational and $U(1)$ symmetries, the symmetry associated with the phason is an \emph{emergent} property of the low-energy theory---one can construct interactions (e.g., three-body interactions) that violate it, but these are irrelevant at low densities and energies. 

We find that there are typically additional phasons in quantum-mechanical (quasi-)crystals, when compared with their classical equivalents. For instance, classically a triangular lattice has no phasons; similarly, for the Penrose quasicrystal ($M = 5$), we find \emph{three} phasons in the quantum-mechanical case, whereas classically only two phasons exist~\cite{bak}. This difference is ultimately due to the fact that the classical order parameters for crystallinity are Fourier components of the \emph{density} (i.e., a real quantum field), whereas in the present case the order parameters are the Fourier components of the microscopic \emph{Bose fields} themselves. Thus, interactions such as $\phi^3$ or $\phi^5$ are forbidden by $U(1)$ symmetry in the quantum case; for classical crystallization, the analogous terms would be allowed. (In the quantum case, the lowest order at which such terms arise is $(\phi^\dagger \phi)^3$ and $(\phi^\dagger \phi)^5$; they are therefore strongly suppressed at low densities.)



\emph{Experimental feasibility}. We now briefly discuss the experimental feasibility of our proposal. First, we note that the Hamiltonian~\eqref{Ham} can be realized in strongly dipolar gases~\cite{dybec, erbiumbec}. For concreteness we consider dysprosium~\cite{dybec}, in which the dipolar interaction naturally exceeds the contact interaction ($R > 2/3$) and quasicrystals can be stabilized. Moreover, dysprosium has a large ground-state manifold, permitting the realization of a nearly symmetric Rashba coupling~\cite{campbell}. We emphasize that a perfectly symmetric spin-orbit coupling is \emph{not} required; nontrivial crystals and quasicrystals can be stable so long as the anisotropy is of order $U(0) n$.

Second, we estimate the achievable transition temperatures. 
As discussed above, the stability criterion permits experiments at relatively high densities (e.g., spacings $\sim 250$ nm), which would boost the achievable condensation temperatures, as well as the barriers ($\sim U(0) n$) between the various ordered states. Following Refs.~\cite{dybec, pfau} let us take all scattering lengths to be $\sim 100 a_0$ (where $a_0$ is the Bohr radius), and  $d_z \sim n^{-1/2} \sim 250$ nm. Then the typical interaction energy scale is $1-5$ nK, which is within the scope of current experiments. 

Finally, as we briefly discuss in the Supplemental Material, nontrivial crystalline states can be realized even outside of the strongly dipolar regime (which might be relevant, e.g., to experiments with chromium~\cite{pfau}), by coupling the bosons to fermions, which mediate attractive RKKY interactions. 

%
%
%
%
%

\emph{Conclusion}. In this Letter we have shown that strongly dipolar Bose gases (such as dysprosium) subject to a Rashba spin-orbit coupling exhibit a variety of nontrivial spatially ordered states, including a pentagonal quasicrystal, hitherto unrealized with ultracold atomic gases. These phases---which can be realized at currently achievable coupling strengths in experiments with highly dipolar bosons---are intrinsically quantum-mechanical. This work paves the way for future explorations of the distinctively quantum-mechanical collective modes, defects, and melting transitions~\cite{berg09, sg:disclin} of quantum quasicrystals. 

\emph{Note added}. While preparing our work for publication we learned of the complementary work of Wilson et al.~\cite{wilson}. While the systems considered are nominally similar, they differ in two crucial respects: (i) we considered a quasi-2D, homogeneous system, whereas Ref.~\cite{wilson} considered a system in a spherical 3D trap; (ii)~we considered large-spin states with predominantly density-density interactions whereas Ref.~\cite{wilson} treated spin-$1/2$ states with Heisenberg interactions. Thus, the phase diagrams in the two cases are quite different. 

\emph{Acknowledgments}. We are grateful to A. Rosch, B. Lev, R. Wilson, C. Henley, and M. Knap for helpful discussions. S.G. acknowledges support from the Harvard Quantum Optics Center. Work performed at Argonne National Laboratory (by I.M.) is supported by the U. S. Department of Energy, Office of Science, Office of Basic Energy Sciences, under Contract No. DE-AC02-06CH11357. E.A.D. acknowledges support from Harvard-MIT CUA, 
the DARPA OLE program, AFOSR MURI on Ultracold Molecules, and ARO-MURI on Atomtronics.

\begin{widetext}

\section{Supplemental Material}

In this document we provide the details of the mean-field analysis leading to the results in the main text. We begin with the ansatz [Eq.~(5) of the main text]:

\beq
\phi(\mathbf{x}) =  \sqrt{ \frac{n}{2M}} \sum_{i = 1}^M \bigg[ e^{i (\mathbf{k}_i \mathbf{\cdot x} + \alpha^+_i)} 
\left( \begin{array}{c} 1 \\ e^{i \theta_i} \end{array} \right) + e^{-i (\mathbf{k}_i \mathbf{\cdot x} + \alpha^-_i)} 
\left( \begin{array}{c} 1 \\ -e^{i \theta_i} \end{array} \right)\bigg],
\eeq
and evaluate its interaction energy 

\beq
\int d^2 q \, \rho(\mathbf{q}) \rho(\mathbf{-q}) U(q)
\eeq
where $\rho_{\mathbf{k}} \equiv \int d^2x e^{i \mathbf{k \cdot x}} (\phi^\dagger_+ (\mathbf{x})\phi_+ (\mathbf{x}) + \phi^\dagger_- (\mathbf{x}) \phi_- (\mathbf{x}))$. 
The first step is to compute the densities $\rho_+(\mathbf{x})$ and $\rho_-(\mathbf{x})$. For general $M$ we find that these are given by

\bea
\rho_+(\mathbf{x}) & = & \frac{1}{2M } \bigg[ M + \sum_{i = 1}^M \cos(2 \mathbf{k_i \cdot x} + \alpha^+_i + \alpha^-_i) + \sum_{\Box} \cos(\mathbf{(k_i - k_j) \cdot x} + \alpha^+_i - \alpha^+_j) \\ && \quad \qquad  + \cos(\mathbf{-(k_i - k_j) \cdot x} - \alpha^-_i + \alpha^-_j) + \cos(\mathbf{(k_i + k_j) \cdot x} + \alpha^+_i + \alpha^-_j) + \cos(\mathbf{-(k_i + k_j) \cdot x} - \alpha^-_i - \alpha^+_j) \bigg]  \nonumber \\
\rho_-(\mathbf{x}) & = & \frac{1}{2M } \bigg[ M - \sum_{i = 1}^M \cos(2 \mathbf{k_i \cdot x} + \alpha^+_i + \alpha^-_i) + \sum_{\Box} \cos(\mathbf{(k_i - k_j) \cdot x} + \alpha^+_i - \alpha^+_j + \theta_i - \theta_j) \\ && \quad \qquad  + \cos(\mathbf{-(k_i - k_j) \cdot x} - \alpha^-_i + \alpha^-_j + \theta_i - \theta_j) - \cos(\mathbf{(k_i + k_j) \cdot x} + \alpha^+_i + \alpha^-_j + \theta_i - \theta_j) \nonumber \\ && \quad \qquad - \cos(\mathbf{-(k_i + k_j) \cdot x} - \alpha^-_i - \alpha^+_j + \theta_i - \theta_j) \bigg]  \nonumber
\eea
Note that the total density is normalized to unity for simplicity. The $\Box$ subscript means that the sum is over distinct rectangles, to avoid double-counting---i.e., each set $(\pm \mathbf{k}_i, \pm \mathbf{k}_{j \neq i})$ is only counted once. Using this notation, we find that the total energy takes the form

\bea
E & = & U(0) + \frac{1}{(2M)^2} \sum_{\Box} U(|\mathbf{k_i - k_j}) \left[2 + \cos(\alpha_{ij}) + \cos(\alpha_{ij} + 2\theta_{ij}) + \cos(\alpha_{ij} + \theta_{ij}) + \cos(\theta_{ij}) \right] \\
& & \qquad \qquad \qquad \qquad + U(|\mathbf{k_i + k_j}|) \left[2 + \cos(\alpha_{ij}) + \cos(\alpha_{ij} + 2 \theta_{ij}) - \cos(\alpha_{ij} + \theta_{ij}) - \cos(\theta_{ij}) \right] \nonumber
\eea
as in the main text (we have used the definitions $\alpha_{ij} \equiv \alpha^+_i - \alpha^-_i - (\alpha^+_j - \alpha^-_j)$ and $\theta_{ij} \equiv \theta_i - \theta_j$). 

This expression can be evaluated for various configurations (Fig.~\ref{energies}), using trigonometry to connect $\theta_{ij}$ to $\mathbf{k_i - k_j}$ and then minimizing the internal (non-geometric) phases $\alpha_{ij}$ for each configuration. The preferred states we find are always symmetric; an example is shown in Fig.~\ref{aniso}. With this in mind, let us turn to the heuristic model discussed in the main text, and adapt the assumption that the interaction has a repulsive core for $k d_z \leq 1$ and an attractive value $-U_\infty$ for larger $k$. In this regime we can simplify the interactions to the form

\beq
E = U(0) - \frac{|U_\infty|}{M^2} \sum_\Box \left(1 + \cos(\alpha_{ij} + \theta_{ij}) \cos(\theta_{ij})\right)
\eeq
which is, of course, minimized for a given $\theta$ by setting $\alpha_{ij} = -\theta_{ij}$. Now let us evaluate the sum. The first ($\theta$-independent) sum is simply $M(M - 1)/2$; one can also see that the second term takes the following form:

\beq
M \!\!\!\!\! \sum_{n \leq (M - 1)/2} \!\!\!\!\! \cos(n \pi/M)
\eeq
Combining these, we get the expression

\beq\label{stab}
E = U(0) - \frac{|U_\infty|}{2M} \left[ (M - 1) + \!\!\!\!\!  \sum_{n \leq (M - 1)/2} \!\!\!\!\! 2 \cos(n \pi/M) \right],
\eeq
so that the energy is optimized by increasing $M$. 

Fig.~\ref{energies} and Fig.~\ref{aniso} show examples of the results of this mean-field analysis. Fig.~\ref{energies} plots the energies of various candidate states as a function of the spin-orbit coupling; Fig.~\ref{aniso} shows that, in the region where a particular symmetric state (in this case, the hexagonal state) has a lower energy than other symmetric states, it is also lower-energy than any asymmetric state. This appears to hold everywhere in the phase diagram, and is expected on intuitive grounds as discussed in the main text.

\begin{figure}[htbp]
\begin{center}
\includegraphics{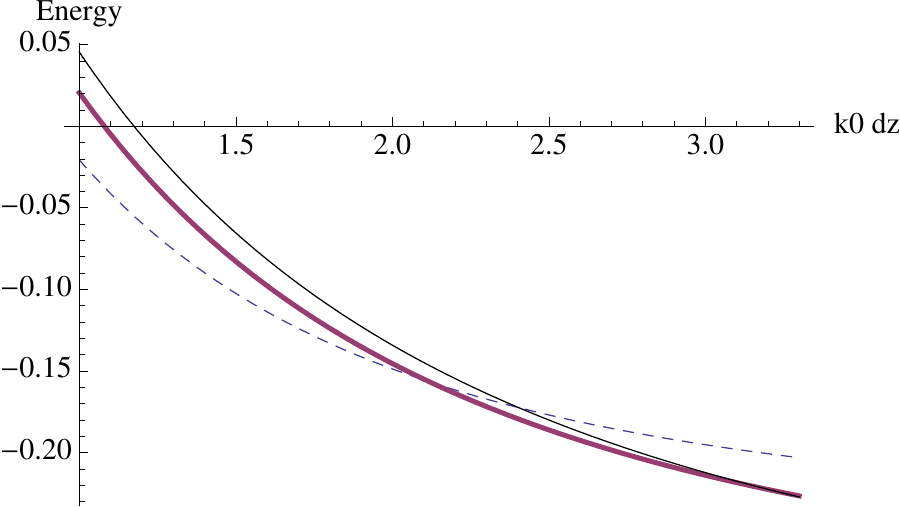}
\caption{Energies of three candidate states---hexagonal (dashed), $M = 5$ quasicrystal (thick), and $M = 7$ quasicrystal (thin)---as a function of spin-orbit coupling $k_0 d_z$, in the dipole-dominated regime $R = 1$.}
\label{energies}
\end{center}
\end{figure}

\begin{figure}[htbp]
\begin{center}
\includegraphics{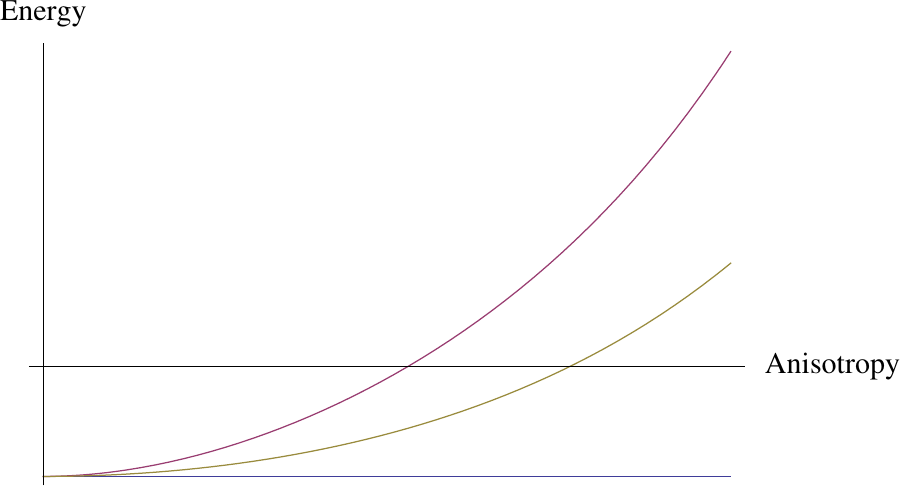}
\caption{Energies of various six-component states as a function of anisotropy. The flat line corresponds to regular hexagons; the other curves correspond to angles $(0, \pi/3 - \theta, 2\pi/3 - \theta)$ [lower] and $(0, \pi/3 - \theta, 2\pi/3 + \theta)$ [upper]. The anisotropic states are higher in energy.}
\label{aniso}
\end{center}
\end{figure}

\subsection{Fermion-mediated interactions}

We now discuss how the nontrivial crystalline phases discussed in the main text can be realized even in systems with weaker dipolar interactions, if the bosons are coupled to a fermionic species. The bosons are taken to interact with the fermions via a symmetric contact interaction

\beq
H_{bf} = V \int d^2x n(\mathbf{x}) \nu_f(\mathbf{x})
\eeq
where $\nu_f$ is the number of fermions; this symmetry is at least approximately satisfied in experiments. The fermions can be integrated out to yield a pure boson-boson interaction of the RKKY form%
\beq
V^{RKKY}_\alpha(q) = - \sum_{\alpha, \mathbf{kk'}} V_{\alpha}^2(q) \chi_\alpha(\mathbf{q}) n_{\mathbf{k}} n_{\mathbf{-k}}
\eeq
where $V_\alpha(q) = V_\alpha$ is constant for a contact interaction, and $\chi(q)$ is the Lindhard function of a free Fermi gas, which has the two-dimensional form
\beq
\chi(q) = \frac{k_F^2}{2\pi E_F} \left\{ 1 - \Theta(q - 2k_F) \sqrt{1 - (2k_F/q)^2} \right\}
\eeq
Thus, the RKKY interaction is momentum-independent for $q \leq 2 k_F$, and does not seem, on its own, to favor any particular crystalline configurations~\cite{solenov}. However, when combined with the dipolar interaction it gives rise to an interaction that is sharply peaked in momentum space, as shown in Fig.~\ref{ubf}. Since the RKKY interaction is always attractive, it can lead to effective boson-boson interactions that are attractive in some regions of momentum space even when $U$ itself is always repulsive.  

\begin{figure}[htbp]
\begin{center}
\includegraphics{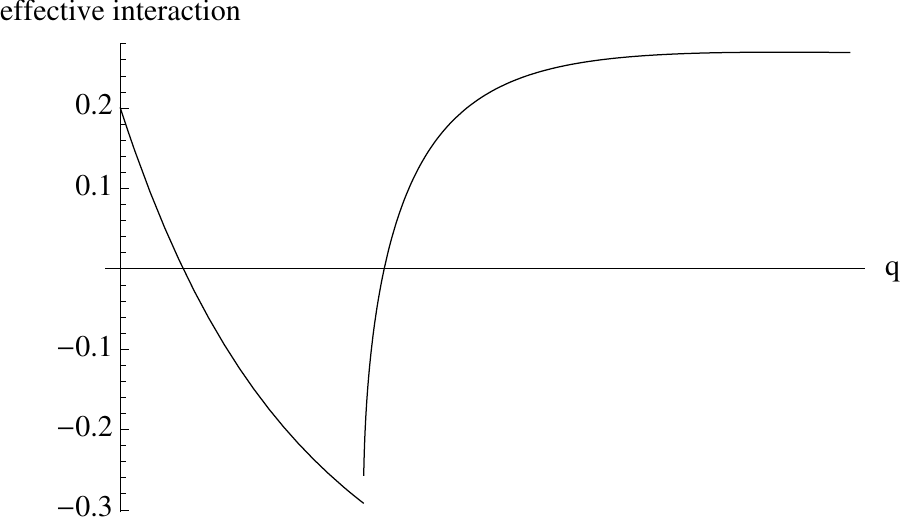}
\caption{Effective bosonic interactions in momentum space due to a combination of dipolar interactions ($R = 0.5$) and RKKY interactions [of strength $0.8 U(0)$].}
\label{ubf}
\end{center}
\end{figure}

\begin{figure}[htbp]
\begin{center}
\includegraphics{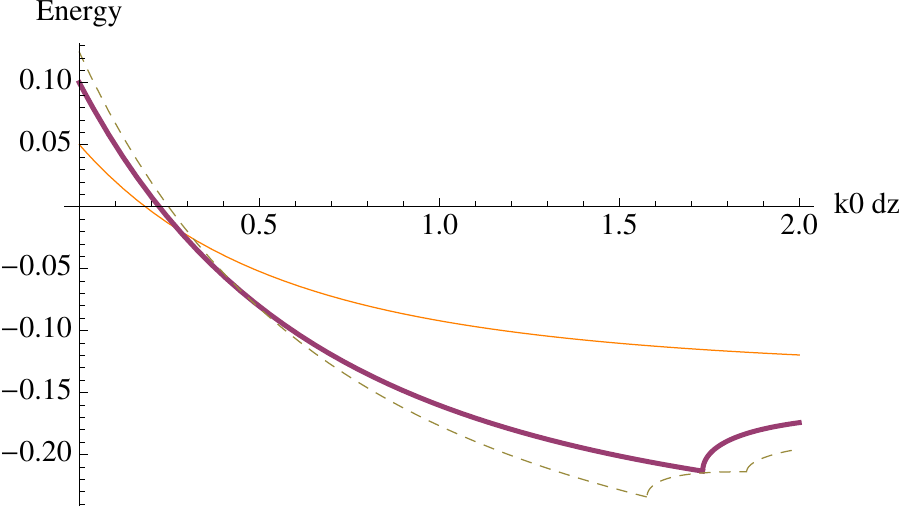}
\caption{Energies of square (dashed), hexagonal (thick) and quasicrystalline (thin) states in the presence of strong RKKY coupling and weak ($R = 0.5$) dipolar interactions. Nontrivial crystalline phases can be stabilized in this regime, but only in the presence of fermions.}
\label{bfenergies}
\end{center}
\end{figure}

\subsection{Stability}

We now fill in a few details of the stability argument adumbrated in the main text. This argument contains two parts: (i)~the system is at least locally stable against spatial collapse (i.e., the interaction energy increases with the density); (ii)~the system is stable against high-momentum Bogoliubov excitations. Statement~(i) can be verified by inspection of Eq.~\eqref{stab}; given stability against global collapse, we can then address statement~(ii) by examining the Bogoliubov spectrum at a fixed density. This analysis is carried out in Ref.~\cite{fischer} in the case without spin-orbit coupling, and can be trivially generalized; the Bogoliubov spectrum takes the following general form [Eq.~(8) of Ref.~\cite{fischer}]:

\beq
E_k \sim \sqrt{k^4 + k^2 m U(k) n}.
\eeq
Evidently, for small enough densities $n$ this equation has no real roots; therefore, the mean-field states discussed in the main text are at least metastable. 

\subsection{Transitions and possible intermediate states}

Finally, we briefly discuss the nature of transitions between the striped state and the various (quasi-)crystalline states. The mean-field analysis in the main text considered only states with equal weight at each of $2M$ momenta. It is clear that the transitions between such states are strongly first order~\footnote{One can see this directly from Fig.~\ref{energies}; a phase transition corresponds to the intersection of two separate curves, and consequently to a discontinuous first derivative of the ground state energy.}. However, in principle one can also imagine symmetry-allowed continuous transitions between the stripe and a crystal. These could arise, for instance, if the stripe became unstable to excitations at a second pair of momenta on the dispersion minimum (Fig.~2(a) of main text). The weight of the Bragg peaks corresponding to this second pair of minima would then (hypothetically) grow continuously from zero. 

We now sketch why such states do not in fact arise, thus justifying our restriction to equal-weight superpositions of condensates. We argue this explicitly for the rectangular lattice ($n = 2$) but the generalization is straightforward. We consider the general wavefunction $\phi(\mathbf{x}) = \sqrt{n} [ A \cos(\mathbf{k \cdot x}) + B \cos(\mathbf{q \cdot x})]$. The interaction energy for this state has the generic form $H_{\mathrm{int}} \sim \mathrm{constant} + \beta A^2 B^2$, where $B^2 = 1 - A^2$. The only possible states that minimize the energy are $A = 1/\sqrt{2}$ (i.e., the fully crystallized state) and $A = 1$ (i.e., the stripe). Thus, we find no intermediate states between the stripe and the equal-weight crystal, and it follows that the phase transition between these two states is expected to be strongly first-order. 

\end{widetext}


\end{document}